\documentclass[reprint,showpacs,amsmath,amssymb,prl,noshowpacs]{revtex4-1}

\usepackage{graphicx} 

\begin{document}

\newcommand{\var}{\mathrm{var}}

\title{Improved detection of small atom numbers through image processing}

\author{C. F. Ockeloen}
\author{A. F. Tauschinsky}
\author{R. J. C. Spreeuw}
\email{R.J.C.Spreeuw@uva.nl}
\author{S. Whitlock}
\affiliation{Van der Waals-Zeeman Institute, University of Amsterdam,\\
Valckenierstraat 65, 1018 XE Amsterdam, The Netherlands}

\date{\today}

\begin{abstract}
We demonstrate improved detection of small trapped atomic ensembles through advanced post-processing and optimal analysis of absorption images. A fringe removal algorithm reduces imaging noise to the fundamental photon-shot-noise level and proves beneficial even in the absence of fringes. A maximum-likelihood estimator is then derived for optimal atom-number estimation and is applied to real experimental data to measure the population differences and intrinsic atom shot-noise between spatially separated ensembles each comprising between 10 and 2000 atoms. The combined techniques improve our signal-to-noise by a factor of 3, to a minimum resolvable population difference of 17 atoms, close to our ultimate detection limit.
\end{abstract}


\maketitle

Trapped ultracold atoms and quantum degenerate gases are novel systems for the study of many-body quantum physics~\cite{BloDalZwerger08} and are key to new technologies such as trapped atom interferometers~\cite{WanAndWu05,*SchHofKruger05,*JoShiPrentiss07} and atomic clocks~\cite{HarLewCornell02,*TreHomReichel04,*AndTicHall09,*DeuRamRosenbusch10}. This is exemplified by recent experiments on number/spin-squeezing and entanglement between small atomic ensembles~\cite{EstGroOberthaler08,*AppWinPolzik09,*RieBTreutlein10,*GroZibOberthaler10,*LerSchVuleti'c10}, which could serve as a resource for quantum metrology and quantum information science. Essential to such applications is the ability to precisely measure the populations of a pair of atomic ensembles in, for example, a double-well potential for Josephson physics or atom interferometry~\cite{WanAndWu05,SchHofKruger05,JoShiPrentiss07,AlbGatOberthaler05,HalWhiSidorov07,LevLahSteinhauer07} or the spin states of an atomic clock. Upon readout the interferometric phase can be mapped to a population difference and the two ensembles are imaged to different regions of a CCD camera. The relative population is robust against common-mode technical fluctuations such as probe noise or trap loading efficiency. However, imaging noise and intrinsic atom number fluctuations (atom shot noise) typically limit measurement precision.

Here we demonstrate improved detection of trapped ensembles of ultracold atoms through advanced post-processing and optimal analysis of laser-illuminated absorption images. This allows us to better measure the intrinsic atom number fluctuations in a magnetic lattice potential. First we apply a fringe-removal algorithm to reduce residual imaging noise to the fundamental photon shot-noise level. We then establish the ultimate limit for measuring the relative populations based on the Cram\'er-Rao bound (CRB) and derive maximum likelihood estimators used to attain this limit~\cite{KayKay98}. These optimal analysis techniques provide the basis to improve the readout of trapped atom interferometers to the quantum limit or to better resolve number squeezing and entanglement.
\begin{figure}%
\includegraphics[width=0.85\columnwidth]{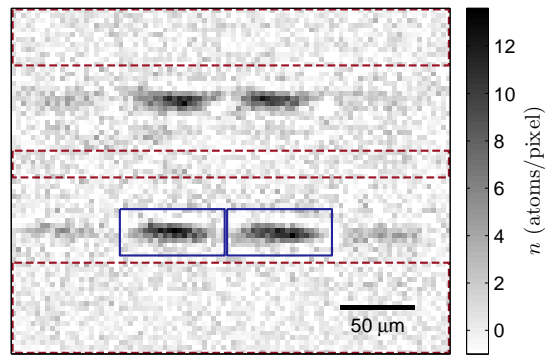}
\caption{Typical absorption image of the atomic distribution ($90\times71$~pixels) for a hold time of $1.6~$s. Four independent atomic ensembles are visible (horizontally distributed) with two mirror images (vertically separated). The two center-most clouds each contain $\approx400$ atoms. Dashed boxes indicate the background pixels for the fringe removal algorithm and the signal regions are indicated by solid lines.}
\label{fig:avgimage}
\end{figure}

In our experiment we prepare up to a total of $N=4\times10^3$ $^{87}$Rb atoms in a multi-well trap formed by a current through a Z-shaped wire and the permanent field of a magnetic lattice atom chip~\cite{GerWhiSpreeuw07,*WhiGerSpreeuw09}. Recently we used this setup to study sub-Poissonian atom number fluctuations resulting from three-body loss in an array of tightly confining microtraps~\cite{WhiOckSpreeuw10}. In this work we position the trap at the edge of the magnetic lattice, $45~\mu$m below the chip surface, to create four potential wells (two deep central wells and two shallow outer wells). The atoms are evaporatively cooled in this trap to a temperature of $\sim1~\mu$K, thereby creating independent atomic ensembles, each occupying an area of $\sim10\times50~\mu$m$^2$ in our images (fig.~\ref{fig:avgimage}). The mean number of atoms is varied by holding the atoms in the trap for 0--4~s at a fixed temperature and partially by reducing the amount of Rb dispensed during loading.

Absorption imaging is the standard method for detecting trapped neutral atoms. Here atoms are briefly exposed to a nearly homogeneous probe laser, typically tuned to resonance with an atomic transition and the resulting absorption signal $A$ is imaged onto a CCD camera. Subsequently the atoms are ejected from the trap and a reference image $R$ is recorded to normalize intensity variations of the probe. A dark image may also be recorded without the probe to subtract any stray light or CCD dark counts. The two-dimensional atomic density is calculated as $n = -\alpha[\log(A/R)+s(A/R-1)]/\sigma_0$~\cite{ReiLahGu'ery-Odelin07}. Here $\sigma_{0}=3\lambda^2/2\pi$ is the absorption cross-section, $\alpha$ is a dimensionless parameter which depends for example on the probe polarization and $s = I_R/\alpha I_s^0$ is the (spatially dependent) saturation parameter, with $I_R$ the probe intensity at the position of the atoms and $I_s^0=1.67~$mW$/$cm$^2$ the saturation intensity for our transition.

The experiments employ a back-illuminated deep-depletion CCD camera with a quantum efficiency of $Q_e\approx0.9$ at $\lambda=780$~nm, a measured gain of $g=0.87\pm 0.05$ counts/photon and a readout noise level of $\sigma_\mathrm{rd}\approx 13$ counts. The resolution of our optical system is $9.6~\mu$m (Rayleigh criterion) with a pixel area of $\Delta = (3.3\pm0.1~\mu\mathrm{m})^2$ in the object plane. The probe is slightly inclined with respect to the gold-coated chip surface to create two mirror images of the atoms (fig.~\ref{fig:avgimage})~\cite{SchKasSchmiedmayer03}. Correlating these images provides additional means to distinguish atom shot noise and imaging noise.

We have experimentally optimized our imaging parameters to provide the highest signal-to-noise ratios (SNR) for in-trap imaging. Long exposure times $\tau$ tend to lower the effect of photon shot noise, however blurring/heating due to photon recoil during the imaging pulse tends to increase the effective area $a$ over which the atoms are distributed; therefore, one finds an optimum intensity and exposure time. We measure the SNR in a series of absorption images for constant atom number as a function of both $\tau$ and $s$ and find a maximum for $\tau=50~\mu$s and $s\approx0.54$. Finally, we calibrate $\alpha$ by comparing the integrated absorption signal with a second set of images taken after free expansion using a weak probe ($s\approx 0.02$), yielding $\alpha = 3.0\pm 0.2$. Detection linearity with respect to atom number variations is confirmed by measuring a fixed ratio between the populations of the center and outer wells. Our typical atom numbers correspond to weak absorption signals ($A/R\gtrsim0.8$) for which it can be difficult to extract a signal buried in noise.

In practice additional imaging noise originates from fringes due to diffraction and interference of the probe beam by optical elements and the atom chip surface. Small vibrations between the absorption and reference images result in imperfect normalization giving rise to fluctuating fringe patterns. This noise can be greatly reduced through the application of a fringe-removal algorithm, while making no assumptions about the atomic distribution. The algorithm works by composing for each absorption image a matching optimal reference image $Q$ constructed as a linear combination of many reference images $R_k$ within a set, $Q = \sum_kc_kR_k$~\cite{fringeremovalref,LiKeWang07}. The method is closely related to that applied to facial recognition~\cite{SirKirKirby87} and recently in astronomical image analysis for detecting extrasolar planets~\cite{LafMarArtigau07}.

To obtain the coefficients $c_k$ we minimize the least squares difference between the absorption and reference images $\sum_xm_x(A_x - Q_x)^2$, where $x$ indexes each pixel, within a specified background region ($m_x=1$) excluding the signal region ($m_x=0$). Setting partial derivatives with respect to $c_j$ to zero, we obtain a linear system of equations, $\sum_k\!c_k B_{j,k}=\sum_xm_x R_{x,j} A_x$ with the square matrix $B_{j,k}=\sum_x\!m_x R_{x,j} R_{x,k}$, which can be readily solved for $c_k$. A typical data set consists of hundreds of absorption images, therefore we decompose $B$ once using LU or singular value decomposition then substitute to obtain $c_k$ for each absorption image~\cite{GolGolub96}. The algorithm is sufficiently fast to decompose $B$ and process new absorption images in $<1$~s for live processing between experimental cycles.

\begin{figure}%
\includegraphics[width=0.9\columnwidth]{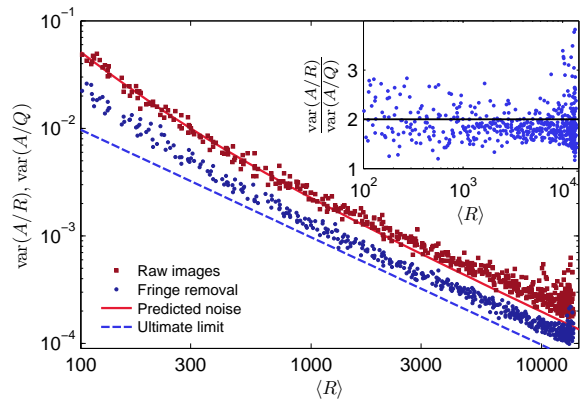}
\caption{Imaging noise for various probe intensities with and without fringe removal, $\var(A/Q)$ (circles) and $\var(A/R)$ (squares) respectively. The predicted photon shot-noise plus camera readout noise is shown as a solid line. The dashed line shows the ultimate photon shot-noise limit.}
\label{fig:photonshotnoise}%
\end{figure}

Figure~\ref{fig:photonshotnoise} shows the imaging noise with and without fringe removal, $\var(A/Q)$ and $\var(A/R)$ respectively, calculated for a signal-free region (separate from the fringe-removal background region) for various probe intensities. Without fringe removal the measured noise is in good agreement with the expected photon shot noise and readout noise given our CCD parameters. Application of the fringe removal algorithm with a basis of $\sim250$ reference images reduces the measured variances by a factor of $1.9\pm 0.3$ over the full range of intensities. Remarkably, even in the absence of fringes, the algorithm reduces the photon shot-noise contribution originating from $R$. This is possible since the optimal reference image is the (weighted) average over many reference images, allowing an additional decrease of uncorrelated noise by up to a factor of $2$. For our chosen imaging parameters the remaining noise from pixel to pixel is $\sigma_n = 1.3\pm 0.3$ atoms, close to the ultimate limit of $1.1$ atoms due to photon shot noise in $A$. There is also a small residual correlated noise component which fluctuates on a length scale comparable to our cloud size with an rms amplitude of $0.06\pm0.01$ atoms.

We wish to measure the total and difference populations $N^\pm=N_p\pm N_q$ and the relative population difference $N^-/N^+$. To establish the limit for extracting these populations from our images we derive the Cram\'er-Rao bound (CRB). The CRB is a powerful tool which gives a lower bound for the variance of any parameter estimate, independent of the exact procedure used to extract the information~\cite{KayKay98}. We describe our image data as the absorption signal from a pair of atomic ensembles corrupted by uncorrelated additive gaussian noise $d_{i,x}$ with variance $\sigma_n^2$: $n_{i,x}=N_{i,p}p_x+N_{i,q}q_x + d_{i,x}$, where $i$ labels a particular realization. The normalized spatial mode functions $p_x$ and $q_x$ ($\sum_xp_x = \sum_xq_x = 1$) are determined by the cloud shape and optical resolution of our imaging system. The log-likelihood function is
\begin{equation}
l(N_{p,q};n_{i,x})\!=\!-\!\sum_x\!\frac{(n_{i,x}\!-\!N_{i,p}p_x\!-\!N_{i,q}q_x)^2}
{2\sigma_n^2}\!+\!\mathrm{const.}
\end{equation}
The CRB is then the inverse of the Fisher information matrix, calculated from the second derivatives of $l$ with respect to $N^\pm$. We obtain for the CRB
\begin{eqnarray}\label{eq:crb}
\textrm{cov}(N^+,N^-)\geq C=\frac{4\sigma_n^2}{(u^+u^-)-v^2}\left(\begin{matrix}
  u^-&-v \\
 -v& u^+
\end{matrix}\right),
\end{eqnarray}
with the parameters $u^\pm=\nobreak \sum_x(p_x\pm q_x)^2$ and $v=\sum_x(p_x^2- q_x^2)$. Here $\var(N^+)\geq C_{11}$, $\var(N^-)\geq C_{22}$, and for the relative atom number,
\begin{equation}\label{eq:varexpansion}
\var\!\left(\frac{N^-}{N^+}\right)\!\geq\!\frac{\langle N^-\rangle^2C_{11}\!+\!\langle N^+\rangle^2 C_{22}\!-\!2\langle N^-\rangle \langle N^+\rangle C_{12}}{\langle N^+\rangle^4}.
\end{equation}
Taking typical numbers from our experiment ($v\approx0$ and $1/u^\pm\approx30$) we find the single-shot CRB to be $C_{11}=C_{22}=200$~atoms$^2$, or a minimum resolvable population difference of $14$~atoms. The CRB is also easily applied to estimating the number of atoms within a single ensemble for which our detection limit is $10$~atoms/shot.

The CRB can be attained by a maximum likelihood estimator (MLE), found by maximizing the log-likelihood function and solving for $N^\pm$,
\begin{equation}\label{eq:mle}
\hat{N}^\pm_i= \frac{2u^\mp}{(u^+u^-)-v^2}\sum_xn_{i,x}\left((p_x\pm q_x)-\frac{v(p_x\mp q_x)}{u^\mp}\right).
\end{equation}
Equation~\eqref{eq:mle} can be interpreted as a sum over the imaged density distribution weighted by $p_x\pm q_x$, where the second term accounts for overlapping and uneven mode functions. This gives a direct measure of the populations while minimizing the influence of noise. The mode functions $p_x$ and $q_x$ can be obtained from the data in a model independent way by averaging over many images to suppress noise. If $p_x$ and $q_x$ are not spatially separated one could record a set of images where each ensemble is individually populated or apply independent component analysis to isolate signal components~\cite{SegDioAnderson10}.

We compare the expected performance of the MLE to naive estimates for $N^\pm$ obtained by separately integrating over rectangular sub-images containing the left and right ensembles. For this case, the expected variance is $\var(N^\pm)\geq a\sigma_\mathrm{n}^2$, with $a$ the total number of pixels in the integration regions. For regions chosen to include $>95\%$ of the atomic signal ($a/2=10\times22$~pixels, fig.~\ref{fig:avgimage}) we expect a detection noise contribution to $\var(N^\pm)$ of 730~atoms$^2$, a factor $3.7$ larger than the CRB.

We have performed measurements of the relative atom number $N^-/N^+$ for varying total atom number to resolve intrinsic atom number fluctuations and demonstrate the improvements gained using maximum-likelihood estimation. Our data set spans 40 hold times between 0--4~s, corresponding to varying $\langle N^+\rangle$ from $4\times 10^3$ down to $\sim20$ atoms. For each hold time we repeat the experiment $40$ times. Sub-pixel image registration is applied to align each image to the average to eliminate small fluctuations in the atom cloud positions. The mode functions are computed by averaging all the data and segmenting the result. We then extract from the images the atom number populations using eq.~\eqref{eq:mle} and compute the variance $\var(N^-/N^+)$. The expected atom shot-noise contribution to the variance, accounting for a mean population imbalance, is $1/\langle N^+\rangle-\langle N^-\rangle^2/\langle N^+\rangle^3$.

\begin{figure}%
\includegraphics[width=0.9\columnwidth]{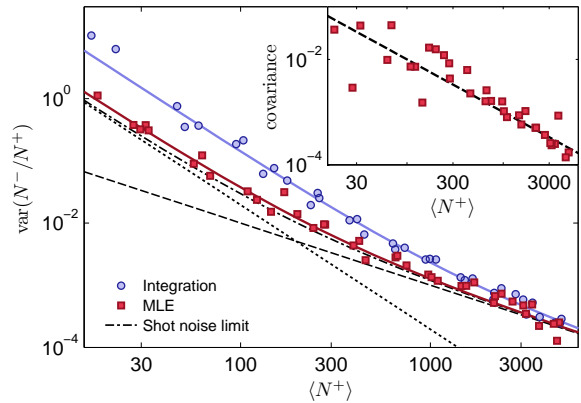}
\caption{Relative variance as a function of total atom number measured by maximum-likelihood estimation (squares) or straight integration (circles). The dash-dotted line shows the expected fluctuations combining atom shot-noise (dashed line) and the minimum detection noise contribution given by the CRB (dotted line). Solid lines indicate fits for the detection noise contributions. The inset shows the covariance between $N^-/N^+$ in the upper and lower mirror image (squares) along with the predicted atom shot noise (dashed line).}
\label{fig:variance}%
\end{figure}

Figure \ref{fig:variance} shows the measured variance of the relative population for both the MLE and simple integration as a function of $\langle N^+\rangle$. We restrict our analysis to the lower mirror image specified in fig.~\ref{fig:avgimage}. For large $\langle N^+\rangle$ both estimation methods yield results which are atom shot-noise limited. For the integration method the measured variance is limited by detection noise for $\langle N^+\rangle$ below $\sim1000$~atoms. Detection is significantly improved using the MLE enabling atom-shot-noise limited detection down to below $300$ atoms. To the data we fit the model $\var(N^-/N^+)= \tilde{C}_{22}/\langle N^+\rangle^2+1/\langle N^+\rangle$ (for our data $\langle N^-\rangle\approx0$), to measure the detection noise contribution. For straight integration we find $\tilde{C}_{22}=(1.3\pm0.2)\times10^3$~atoms$^2$, whereas for the MLE $\tilde{C}_{22}=270\pm40$~atoms$^2$, corresponding to a minimum resolvable difference of $17$~atoms/shot. The measured noise is slightly larger than the minimum value given by the CRB, but is consistent with the added effect of the residual correlated noise.

We can further distinguish atom number fluctuations from detection noise in our data by comparing the measured population differences in the two mirror images. Here we expect the detection noise to be uncorrelated, and true atom number fluctuations to be correlated between the two images. To verify this, we measure $N^+$ and $N^-$ for both mirror images separately and calculate the covariance of $N^-/N^+$ between the upper and lower images for each hold time. The result is shown in the inset of fig. \ref{fig:variance}, along with the expected atom shot-noise, $1/\langle N^+\rangle$. Our data is consistent with atom shot noise over the full range of atom numbers, demonstrating a robust method to observe intrinsic atom number fluctuations.

To summarize, we have demonstrated improved detection of small atomic ensembles, to close to the ultimate photon shot-noise limit. A fringe removal algorithm and maximum likelihood estimation are applied to absorption images to reach a measured detection sensitivity of $17$ atoms/shot for population differences. The measured variance of the relative populations is in excellent agreement with the lower limit given by atom shot noise and the Cram\'er-Rao bound for our imaging parameters. Combining fringe removal and maximum likelihood estimation we have improved our signal-to-noise ratio by a factor of $3$ allowing for atom shot-noise limited detection of ensembles comprising as few as $270$ atoms, a factor $9$ lower than without these methods. Averaging measurements from both mirror images would offer a further $\sqrt{2}$ improvement in SNR.

The sensitivity could be improved to the single-atom regime by increasing the imaging resolution and better localizing the atoms. Optimal imaging parameters are found by considering a model for blurring due to photon recoil after release from the trap, where the cloud size $a$ increases proportional to $s\tau^3/(1+s)$. Integrated photon shot-noise scales with $\sqrt{a(1+s)^2/s\tau}$, yielding an optimal saturation parameter of $s_\mathrm{opt}=2/3$. For realistic imaging parameters (optical resolution of $1.2~\mu$m, $Q_e=0.9$ and $\alpha=1$) we find an optimal exposure time of $13~\mu$s and a detection limit of $0.5$ atoms/shot for a single ensemble. Further improvements would be possible using squeezed light to image ultracold atoms~\cite{DelTreR'efr'egier08}.

The atom clouds we prepare are considerably elongated, highlighting the potential application of our analysis to the study of 1D quantum gases. Applying maximum likelihood estimation column-by-column provides a robust measure of linear density. For our experimental conditions we infer a sensitivity of $0.8$ atoms/$\mu$m. This is below the typical linear density required to reach the crossover from the weakly interacting to the strongly interacting regime on an atom chip~\cite{Olshanii98,*ReiThywissen04,*EsWicDruten09}, which could be directly imaged in a single realization. By applying a weak optical lattice along the length of the cloud, it would be possible to directly observe the 1D Mott insulator phase transition~\cite{BloDalZwerger08,HalHarNagerl10}, with a sensitivity of a single atom per lattice site.

\begin{acknowledgments}
We would like to thank N. J. van Druten for loan of the CCD camera and fruitful discussions. We are grateful to FOM and NWO for financial support. SW acknowledges support from a Marie-Curie fellowship (PIIF-GA-2008-220794).
\end{acknowledgments}

\bibliography{cockeloen10}

\end{document}